\documentclass[12pt]{article}
\usepackage{epsfig}
\newcommand{\beq}{\begin{equation}}
\newcommand{\eeq}{\end{equation}}
\newcommand{\beqa}{\begin{eqnarray}}
\newcommand{\eeqa}{\end{eqnarray}}
\newcommand{\beqar}{\begin{eqnarray*}}
\newcommand{\eeqar}{\end{eqnarray*}}

\newcommand{\norm}[1]{\raise.3ex\hbox{:}#1\raise.3ex\hbox{:}}

\begin{document}

\setlength{\unitlength}{1mm}

\thispagestyle{empty}
\rightline{\small gr-qc/0310135 
\newline \small \hfill FREIBURG-THEP-03-16}
\vspace*{3cm}

\begin{center}
{\bf \Large Rotating Boson Stars in $(2+1)$ Dimensions}\\
\vspace*{2cm}

{\bf Dumitru Astefanesei}\footnote{E-mail: 
{\tt astefand@hep.physics.mcgill.ca}}
{\bf and Eugen Radu}\footnote{E-mail: 
{\tt radu@newton.physik.uni-freiburg.de}}

\vspace*{0.2cm}

{\it $^{1}$Perimeter Institute for Theoretical Physics}\\
{\it 35 King Street North, Waterloo, Ontario N2J 2W9, Canada}\\[.5em]

{\it $^{1}$Department of Physics, McGill University}\\
{\it Montr\' eal, Qu\' ebec H3A 2T8, Canada}\\[.5em]

{\it $^2$Physikalisches Institut, Albert-Ludwigs-Universit\"at 
Freiburg} \\
{\it Hermann-Herder-Stra\ss e 3, D-79104 Freiburg, Germany}

\vspace{2cm} {\bf ABSTRACT} 
\end{center}
We consider rotating boson star solutions in a three-dimensional 
anti-de 
Sitter spacetime and investigate the influence of the rotation on 
their
properties.  The mass and angular momentum of these configurations 
are computed by using the counterterm method.
No regular solution is found in 
the limit of vanishing cosmological constant.
\vfill \setcounter{page}{0} \setcounter{footnote}{0}
\newpage
\section{Introduction}

The problem of finding globally regular soliton solutions  with a 
nonvanishing angular momentum
has recently enjoyed a renewed interest, various physical systems 
having been considered in the literature.
A somewhat unexpected result obtained in this context was the absence 
of
rotating regular solutions with a nonvanishing magnetic charge  
in a spontaneously broken nonabelian gauge theory 
\cite{Volkov:2003ew}-\cite{vanderBij:2002sq}.

To our knowledge, the only explicit example of 
rotating solitons in asymptotically flat space are found in a complex 
scalar field theory.\footnote{Note also the existence of 
slowly rotating Einstein-Yang-Mills solitons \cite{Brodbeck:1997ek}; 
however, they have been found within a nonperturbative approach in an
anti-de Sitter (AdS) spacetime only \cite{Radu:2002rv}, 
while their existence in asymptotically flat space 
is unclear \cite{VanderBij:2001nm,Kleihaus:2002ee}.}
Within general relativity, the properties of the corresponding 
rotating boson star (BS) 
solutions are discussed in 
refs.~\cite{Yoshida:1997qf}-\cite{Ryan:1996nk}.
BS are well-known gravitational bound states of complex scalar 
fields, providing us with
the simplest model of relativistic stars.
These objects were first studied by Kaup  \cite{Kaup} 
as well as Ruffini and Bonazzalo  \cite{Ruffini:1969qy}, and, 
since then, a large number of papers 
have been published on this subject, 
including a number of reviews (see $e.g.$ \cite{Jetzer:1992jr}-\cite{cqg}).
Ignoring the effects of gravity, the analogous of BS are Q-balls. 
These are 
soliton solutions for a complex scalar field with a 
non-renormalizable 
self-interaction. 
Four dimensional, spinning Q-ball solutions have been recently 
constructed in ref.~\cite{Volkov:2002aj}.

As usual, it is of interest to see how the dimensionality $D$
of spacetime affects 
the properties of these rotating solutions. 
The first obvious 
case is $D=3$, where the problem simplifies dramatically to 
solving a set of ordinary differential equations. 
Also, it would be desirable to have available a lower-dimensional 
toy-model which could exhibit the key features without unnecessary 
complication.
Gravity in three dimensions has attracted much attention in recent 
years, since 
Ba\~nados, Teitelboim and Zanelli (BTZ) found a black-hole spacetime 
\cite{Banados:wn}, 
which provides an important testing ground for quantum gravity and 
AdS/CFT correspondence.
Many other types of 3D solutions have also been found by coupling 
matter fields 
to gravity in different ways.
Nongravitating, spinning Q-balls in a (2+1)-dimensional flat 
background have been 
constructed in ref.~\cite{Volkov:2002aj}, and present very similar 
properties to their four-dimensional counterparts (see also ref. \cite{kim1}).

In this letter we address the problem of finding 
rotating BS  solutions in a three dimensional 
spacetime.\footnote{Rotating stars in a 
(2+1)-dimensional AdS spacetime 
are discussed in refs.~\cite{Lubo:1998ue,Cruz:1994ar},
and present interesting properties. 
However, the matter sources of these configurations 
do not have a field theory interpretation.}
This is rather special case, since, in a 
remarkable development, exact BS static solutions with a negative 
cosmological
constant, $\Lambda<0$, were found in ref.~\cite{Sakamoto:1998hq},
in the limit of large self-interaction (see also 
ref.~\cite{Sakamoto:1998aj}).
Working in the same limit, rotating BS solutions with  $\Lambda<0$
have been found numerically  in ref.~\cite{Sakamoto:1999zt}.
However, the large self-interaction limit considered in 
ref.~\cite{Sakamoto:1999zt} makes obscure the influence 
of a number of physical parameters on the solutions' 
properties, without significantly simplifying the field equations.

Static, circularly-symmetric, non self-interacting BS solutions
of the three-dimensional gravity with negative cosmological constant
were discussed in a more general context in 
ref.~\cite{Astefanesei:qy}.
Similar to the well-known spherically-symmetric four-dimensional 
case,
these circularly symmetric configurations comprise a two-parameter 
family,
labeled by $(\phi_0,n)$, where $\phi_0$ is the central value of the
scalar field and $n$ is the node number $n=0,1,\dots$ of the scalar 
field.
However, there are also major differences, $e.g.$ the existence in 
three dimensions
of a maximal allowed value for $\phi_0$ and 
the absence of local extrema for particle number and total mass. 
Also, no regular solutions are found in the asymptotically flat 
limit.
Here we generalize these static solutions by including a rotating 
term in the general ansatz.

\section{The ansatz and general relations}
We consider a complex scalar field $\Phi$  with a potential $V(|\Phi|^2)$
minimally coupled 
to AdS gravity. The corresponding action of the system is
\begin{eqnarray}
\label{action}
S=-\int_{\mathcal{M}} d^3 x  \sqrt{-g}  
\left(
\frac{1}{16\pi G }(R-2\Lambda) 
- ( g^{i j}\Phi,_{i}^{\ast} \Phi,_{j}
+ V(|\Phi|^2) 
\right)
+\frac{1}{8\pi G }\int_{\partial\mathcal{M}} d^{2}x\sqrt{-h}K,
\end{eqnarray}
where the second term is the Hawking--Gibbons surface term 
\cite{Gibbons:1976ue}. Here, $K$ is the trace of the extrinsic 
curvature of the boundary ($\partial\mathcal{M}$ at spatial infinity) 
and $h_{ab}$ is the induced metric on the boundary, 
while the asterisk denotes complex conjugation. Throughout this letter we
set $c=\hbar=1$; also, the indices $\{i,j,...\}$ will indicate the bulk 
coordinates and $\{a,b,...\}$ will indicate the intrinsic coordinates of 
the boundary metric.
The field equations are obtained by varying the action (\ref{action})  
with respect 
to the field variables $g_{ij}$ and $\Phi$. They are
\begin{eqnarray}
\label{Einstein-eqs}
R_{ij}-\frac{1}{2}g_{ij}R+\Lambda g_{ij}&=&8\pi G  T_{ij},
\\
\label{KG-eqs}
\left(\nabla^2-\frac{dV}{d|\Phi|^2}\right)\Phi &=&0,
\end{eqnarray}
where the energy momentum tensor is defined by
\begin{eqnarray}
\label{Tij}
T_{ij}=\Phi^{\ast}_{,i}\Phi_{,j}+\Phi^{\ast}_{,j}\Phi_{,i}
-g_{ij}(g^{km}\Phi^{\ast}_{,k}\Phi_{,m}+V(|\Phi|^2)).
\end{eqnarray}
Since the theory possesses a global $U(1)$ symmetry, there is a 
conserved 
Noether current
\begin{eqnarray}
\label{J1}
J^{k}=i g^{kl}
\left (\Phi^{\ast}_{,l}\Phi - \Phi_{,l}\Phi^{\ast} \right),
\end{eqnarray}
and an associated conserved charge, namely, the number of scalar 
particles
\begin{eqnarray}
N=\int d^{2} x \sqrt{-g} J^t.
\end{eqnarray}
Working in (2+1) dimensions, we consider a metric ansatz of the form
\begin{eqnarray}
\label{metric}
ds^{2}=\frac{dr^2}{F(r)}+r^{2}(d \varphi +\Omega(r) d 
t)^2-F(r)e^{-2\delta(r)}dt^2,
\end{eqnarray}
where, following the $D$-dimensional ansatz used in 
ref.~\cite{Astefanesei:qy}, we take
$
F(r)=1- 2m(r)+r^2/l^2
$ (with $\Lambda=-1/l^2$).
The coordinate range is $0 \leq r <\infty,~-\infty <t<\infty$,
while the angular variable $\varphi$ is assumed to vary between $0$ 
and $2 \pi$.

The scalar field ansatz considered here is $\Phi=\phi(r) 
e^{i(k\varphi-\omega t)}$, where 
$\phi(r)$ is a real function and $\omega$ is a real constant.
The uniqueness of the scalar field under 
a complete rotation $\Phi(\varphi)=\Phi(\varphi+2\pi)$ 
requires $k$ to be an integer,
which we will call the vorticity number from now on.
Models with $k=0$ corresponds to static, circularly-symmetric 
configurations discussed 
in ref.~\cite{Astefanesei:qy}.
The expression for the particle number is
\begin{eqnarray}
\label{number}
N=4 \pi
\int_{0}^{\infty}dr~ \phi^2 r \frac{e^{\delta}}{F}(\omega+k\Omega).
\end{eqnarray}
We also mention the following relation between the total angular 
momentum $\bar{J}$ and the particle number
\begin{eqnarray}
\label{JN}
\bar{J}=kN,
\end{eqnarray}
which is valid for Kerr-like rotating BS in 
any dimension $D\geq 3$ and $\Lambda\leq 0$.
The above relation has originally been found for four-dimensional asymptotically flat 
BS \cite{Schunck} and holds 
also for Q-ball configurations \cite{Volkov:2002aj}. 
The angular momentum was defined here as 
\begin{eqnarray}
\label{J}
\bar{J}=\frac{1}{8 \pi G}\int 
R_{i}^{j}\xi^{i}_{(\varphi)}d^2\Sigma_{j}=
\int dr d\varphi \sqrt{-g}T_{\varphi}^t = 
4k  \pi\int_0^{\infty} \phi^2 \frac{e^{2\delta}}{F}(\omega+\Omega k),
\end{eqnarray}
and agrees with the definition obtained later by using the 
counterterm 
prescription. Thus, the angular momentum is quantized, which clearly 
contrasts with the rotating perfect fluid star solutions discussed in 
ref.~\cite{Lubo:1998ue}.

The results presented in this letter correspond to a simple scalar
potential $V(\phi)=\mu^2 \phi^2$, (where $\mu$ is the scalar field mass),
although the field equations have been integrated 
also for more complicated forms of $V$.

Since it is convenient to use dimensionless quantities in numerical 
computation, 
we perform the rescalings $r \to r/\mu $,
$\phi \to  \phi/\sqrt{ 16\pi G },~\Lambda \to \Lambda/\mu^2,
~\Omega \to \Omega \omega$, 
while the factor $\omega/\mu$ is absorbed into the definition
of the metric function $\delta$.
With these conventions, we find the field equations
\begin{eqnarray}
\label{eqm}
m'&=&\frac{1}{4}e^{2\delta}r^3\Omega'^2+\frac{1}{2}rF \phi'^2+
\frac{1}{2r}k^2\phi^2+\frac{1}{2}r V(\phi)
+\frac{1}{2F}e^{2\delta}r(1+k\Omega)^2\phi^2,
\\
\label{eqdelta}
(e^{-\delta})'&=&r\left(e^{-\delta}\phi'^2+ e^{\delta}(1+k\Omega)^2\frac{\phi^2}{F^2} \right),
\\
\label{eqfi}
(re^{-\delta}F\phi')'&=& \frac{1}{2}r e^{-\delta}\frac{\partial V }{\partial \phi}
+e^{-\delta}k^2\frac{\phi}{r}
-e^{\delta}\frac{r}{F}(1+k\Omega)^2\phi,
\\
\label{eqO}
(r^3e^{\delta} \Omega')'&=&2ke^{\delta}\frac{r}{F}(1+k\Omega)\phi^2,
\end{eqnarray}
where a prime denotes the derivative with respect to $r$,
while the expression for the energy density is 
\begin{eqnarray}
\label{ro}
\rho=-T_t^t=(1-\Omega^2k^2)e^{2\delta}\frac{\phi^2}{F}+V(\phi)+
\frac{k^2\phi^2}{r^2}+\phi'^2F.
\end{eqnarray}
For nonsingular solutions, the boundary conditions at the origin 
should be $k\phi(0)=0$,
so the scalar field must vanish at the origin for rotating 
configurations,
in contrast with the static $k=0$ case.
At spatial infinity, the scalar 
field must vanish and the metric approaches a BTZ form.
If we assume that this metric possesses a symmetry center
(located at $r=0$), and has no conical singularities,
we have to impose the condition $\lim_{r \to 0} m(r)=0$ 
\cite{Lubo:1998ue}, 
while the function $\Omega(r)$ can 
be nonzero in the same limit.

For small $r$, a power series solution gives for $k\neq 0$ (the 
corresponding 
expansion for static configurations is given in 
ref.~\cite{Astefanesei:qy})
\begin{eqnarray}
\label{expansion}
\nonumber
\phi(r)&=&cr^{|k|}+O(r^{|k|+1}),
\\
\nonumber
\label{expansion-m}
\Omega(r)&= 
&\Omega_0+\frac{c^2|k|(k\Omega_0+1)}{2k(|k|+1)}r^{2|k|}+O(r^{2|k|+1}),
\\
\delta(r)&=&\delta_0-\frac{1}{2} c^2 |k| r^{2|k|}+O(r^{2|k|+1}),
\\
\nonumber
m(r)&=&\frac{c^2|k|}{2}r^{2|k|}+O(r^{2|k|+1}),
\end{eqnarray}
where $c$ is a constant (it is enough to consider $c>0$ only), 
while $\delta_0$ and $\Omega_0$ are 
determined 
by the behavior in the asymptotic region. $\Omega_0$ corresponds
to the angular velocity of the star near its center of rotation.

The analysis of the field equations as $r\to\infty$ gives
\begin{eqnarray}
\label{asymptotics}
\nonumber
\phi(r)&\sim &\hat{\phi}_0 r^{\alpha}+O( r^{\alpha-1}),
\\
m(r)&\sim & M-\frac{1}{2}\frac{J^2}{r^2}
+\frac{ \hat{\phi}_0^2 }{4(\alpha+1)}
(1+\frac{\alpha^2}{l^2})r^{2(\alpha+1)}+O( r^{2\alpha+1}),
\\
\nonumber
e^{-\delta(r)} &\sim & 1+ 
\frac{\hat{\phi}_0^2\alpha}{2}r^{2\alpha}+O( r^{2\alpha-1}),
\\
\nonumber
\Omega &\sim & \frac{J}{r^2}+\frac{k l^2 \hat{\phi}_0^2}{2\alpha 
(\alpha-1)}r^{2(\alpha-1)}+O( r^{2\alpha-3}),
\end{eqnarray}
where $\alpha=-1-\sqrt{1+l^2}$ and $M,\hat{\phi}_0$ and $J$ are 
constants.
We note that, as in other physical situations involving a 
massive scalar field \cite{VanderBij:2001ah},
a nonzero cosmological constant implies a complicated power decay at 
infinity,
rather than an exponential one, as is found in
an asymptotically flat space.
However, the asymptotic behavior of the metric is truly
AdS ($i.e.$ $|g_{tt}| \sim  r^2/l^2 +O(r^0)$, without linear terms in 
$r$).
Therefore, as shown in ref.~\cite{Brown:nw}, the asymptotic symmetry 
group
is the conformal one, which contains the AdS group as a subgroup.

\section{Numerical results and properties of solutions}

So far, we have not found an exact solution of the equations of 
motion even in the absence of rotation, and the resulting system had, 
instead, to  be solved numerically. 
Numerical arguments 
\newpage
\setlength{\unitlength}{1cm}

\begin{picture}(18,8)
\centering
\put(2,0.0){\epsfig{file=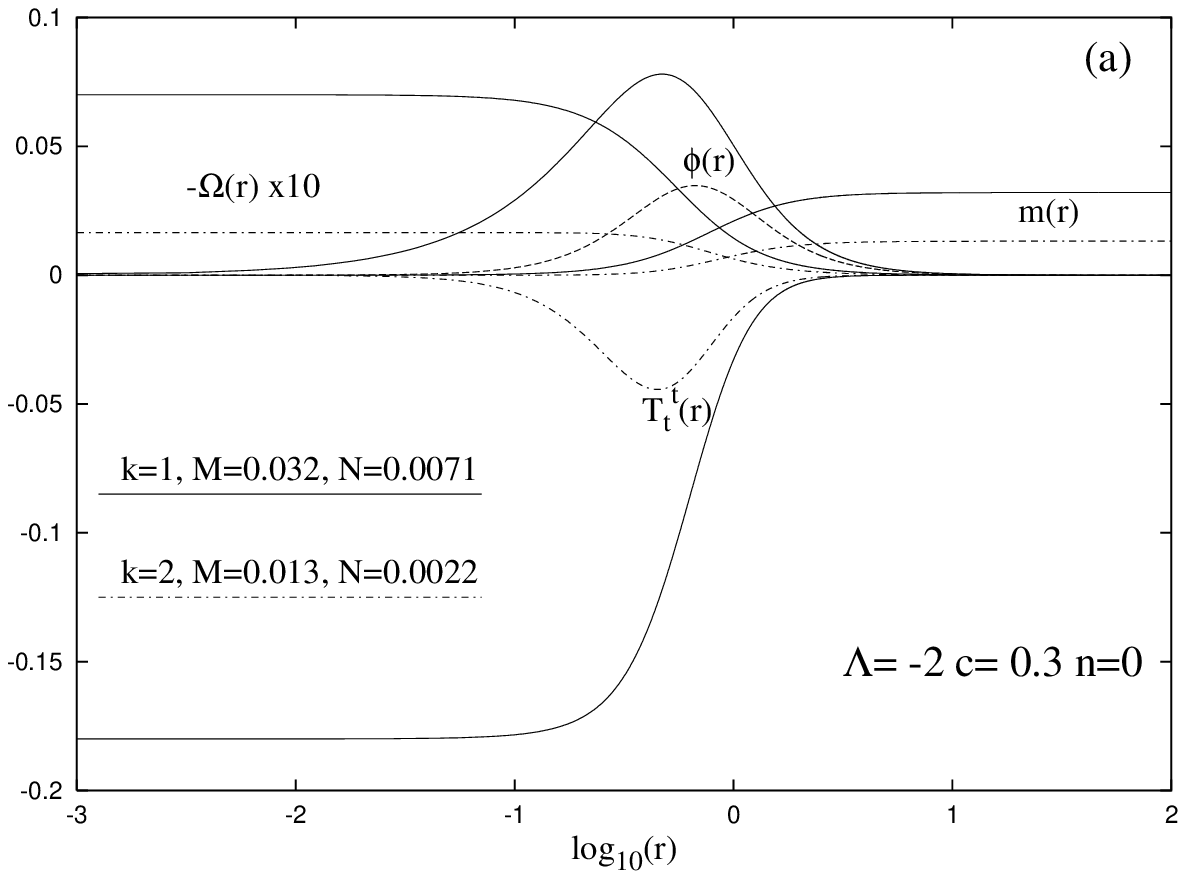,width=12cm}}
\end{picture}
\begin{picture}(19,8.5)
\centering
\put(2.6,0.0){\epsfig{file=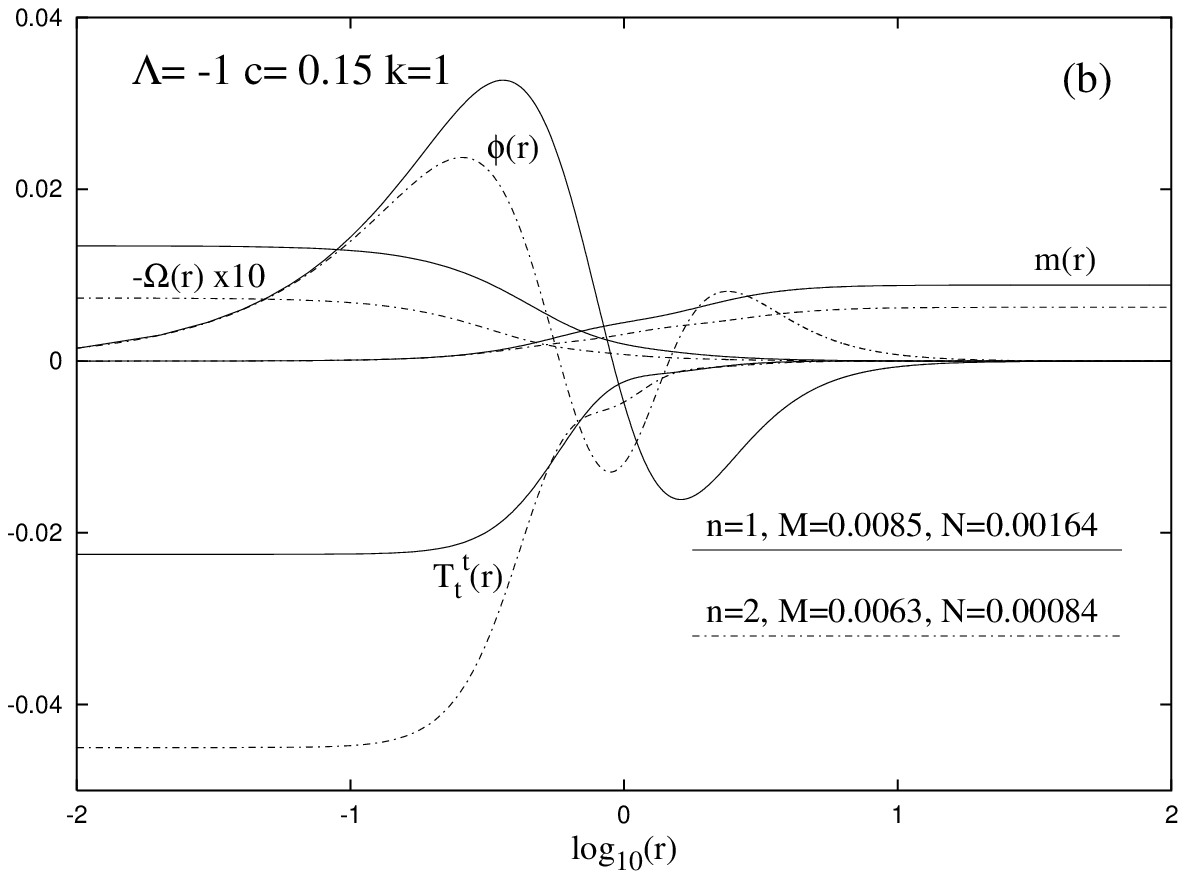,width=12cm}}
\end{picture}
\\
\\
{\small {\bf Figure 1.} The metric functions $m(r),\Omega(r)$, the scalar field $\phi(r)$ and
and the energy density
are shown as a function of the radial coordinate
$r$ for typical solutions
with node number $n=0$ and azimuthal number $k=1,2$ (Figure 1a).
Higher node solutions with $k=1$  are presented in Figure 1b.
Here and in Figure 2, the particle number $N$ is given in units $1/(4G \mu)$.}
\vspace{0.7cm} 
\\
for the existence of $k=0$ 
static, non self-interacting BS solutions 
are presented in ref.~\cite{Astefanesei:qy}.
Here, we will investigate models with nonzero values of the vorticity number
(since the field equations are invariant under $k \to -k,~\Omega \to -\Omega$ 
it is enough to consider positive values of $k$ only).

Following the usual approach, and by using a standard ordinary  
differential  
equation solver, we
evaluate  the  initial conditions (\ref{expansion}) at  $r=10^{-5}$ 
for  global  
tolerance  
$10^{-12}$, adjusting  for fixed shooting parameters and integrating  
towards  $r\to\infty$.
We have found that, given ($c,k, \Lambda$), 
solutions may exist for a discrete set of shooting parameters
 $(\delta_0, \Omega_0)$.
Different values of $(\delta_0, \Omega_0)$
correspond to different numbers, $n$, of nodes of the scalar field.
To simplify the general picture, we consider here nodeless 
configurations mainly.
However, we investigated solutions  with up to four nodes and found 
that they possess a similar behavior to $n=0$ configurations.
Also, the field equations (\ref{eqm})-(\ref{eqO}) have been 
integrated
for values of the cosmological constant  $0<|\Lambda|<100$ and 
vorticity numbers up to ten, finding always the same qualitative 
picture.
The profiles of the functions $m,\phi$ and $\Omega$ and the energy 
density $T_t^t$ 
for typical rotating solutions with the same values of $c$ and 
$\Lambda$ are given 
in Fig.~1. 

We can see that, for given $c,\Lambda$ and $n$, the asymptotic value 
$M$ of the metric function
$m(r)$, and the particle number $N$, decrease 
with the vorticity number. 
Also, unexpectedly, both $M$ and $N$ decrease as the node number is 
increased, for the same values of $c,\Lambda$ and $k$.
The metric functions $m(r)$ and $e^{-\delta(r)}$  
are monotonically increasing, since the right-hand side of 
eqs.~(\ref{eqm}) and (\ref{eqdelta}) is always nonnegative.
Due to the anisotropy of the stress energy tensor, the configurations 
are 
differentially rotating, the rotation function $\Omega(r)$ starting 
with
a nonvanishing value at the origin and monotonically decreasing to 
zero at infinity.
For all solutions we have considered, the metric functions
are completely regular and show no sign of an apparent horizon. 
Also we found no ergoregion, and obviously, no causal anomalies,
usually associated with rotation.

For given $\Lambda, n$ and $k$, we find nontrivial solutions 
up to a maximal value of $c$, where the numerical iteration 
diverges.
The value of $c_{max}$ increases with $|\Lambda|$, while 
the solution with $c=0$ corresponds to the global $AdS_3$.
The numerical integration results for $M$ and $N$ are presented 
in Fig.~2(a) as a function of $c$, for three distinct values of 
$\Lambda$
(note the absence of local extrema for $M$ and $N$).  
The variation of $M$ and $N$ with $\Lambda$, for fixed values of 
the parameter $c$, is shown in Fig.~2(b).
The energy of solutions with the same $c$ decreases with $|\Lambda|$ 
and a
divergent result is obtained in the limit $\Lambda \to 0$.
This is an expected result, since a similar behavior 
has been found for static solutions \cite{Astefanesei:qy}.
It can easily be proven that the rotating term does not affect the 
nonexistence result found in ref.~\cite{Astefanesei:qy} for BS in 
asymptotically flat (2+1)-dimensional spacetime.\footnote{
This result can be viewed as a consequence of the 
absence of self-interaction terms in the scalar field potential.
It can be proven that the nonexistence theorem presented in \cite{Astefanesei:qy}
is not valid if $V$ satisfies 
the condition 
$ \phi  \partial V/\partial \phi -2V <0$.
The field equations have been solved also for a scalar potential on the form 
$V =\mu^2 \phi^2 (1+\lambda_1\phi^2+\lambda_2\phi^4)$.
For $\lambda_1^2 <4\lambda_2$,
this scalar potential is known to present flat space nongravitating solutions
\cite{Volkov:2002aj, kim1}. 
However, the qualitative properties of the solutions we found 
are rather similar to the non-selfinteracting case.
In particular we failed again to find solutions in the limit $\Lambda \to 0$.
}
Also, we have every reason to believe that the no hair theorem 
forbidding the
existence of static black hole solutions with a harmonically
time-dependent scalar field \cite{Astefanesei:qy} 
can be generalized for rotating configurations.

The value of $\rho=-T_t^t$ at the origin is nonzero only 
for $k=0,\pm 1$ configurations, and vanishes for other vorticity numbers.
The energy density is concentrated in an effective mass-circle 
(for $k=0, \pm 1$), or in a ring shape, for other values of $k$ (see 
Fig.~1).
Thus, for $|k|>1$, this situation 
\newpage
\setlength{\unitlength}{1cm}

\begin{picture}(18,8)
\centering
\put(2,-.7){\epsfig{file=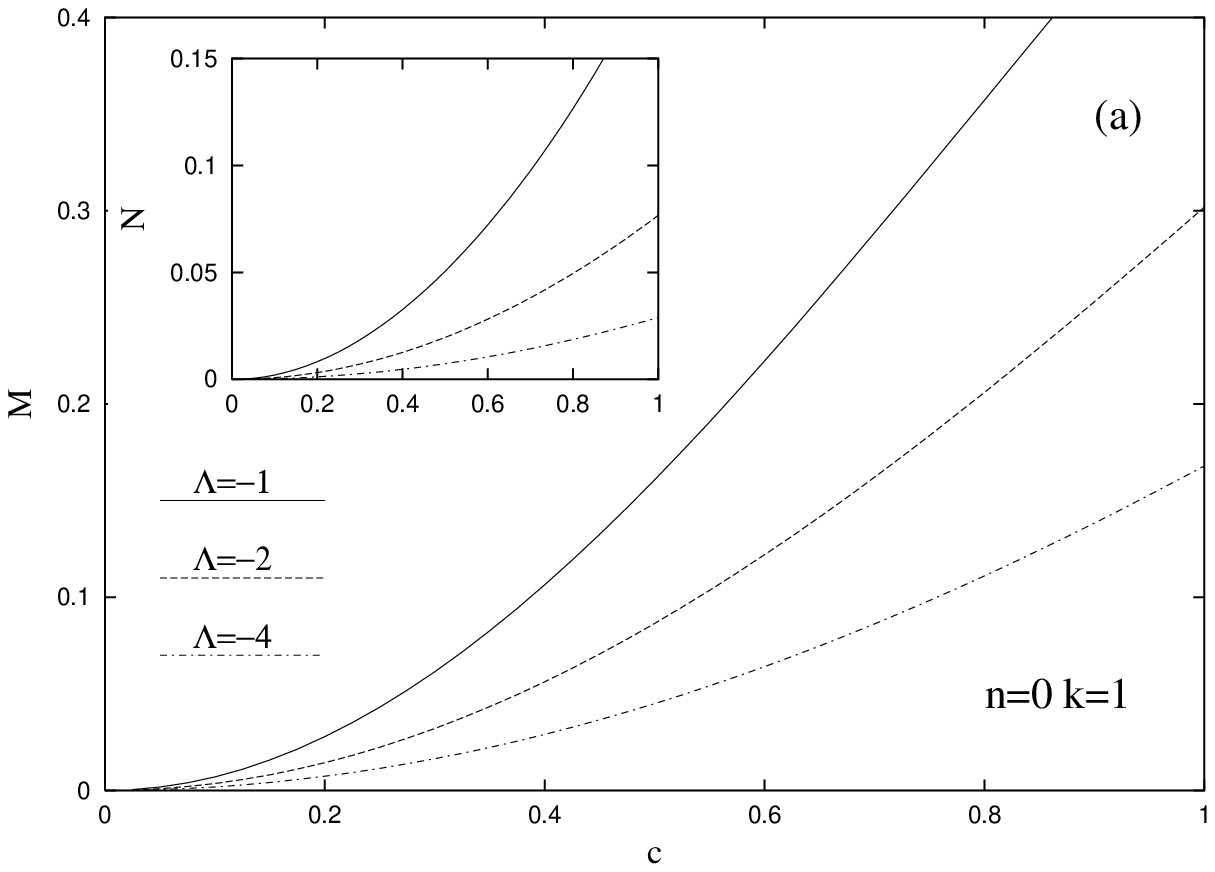,width=13cm}}
\end{picture}
\begin{picture}(19,8.5)
\centering
\put(2.6,-1.1){\epsfig{file=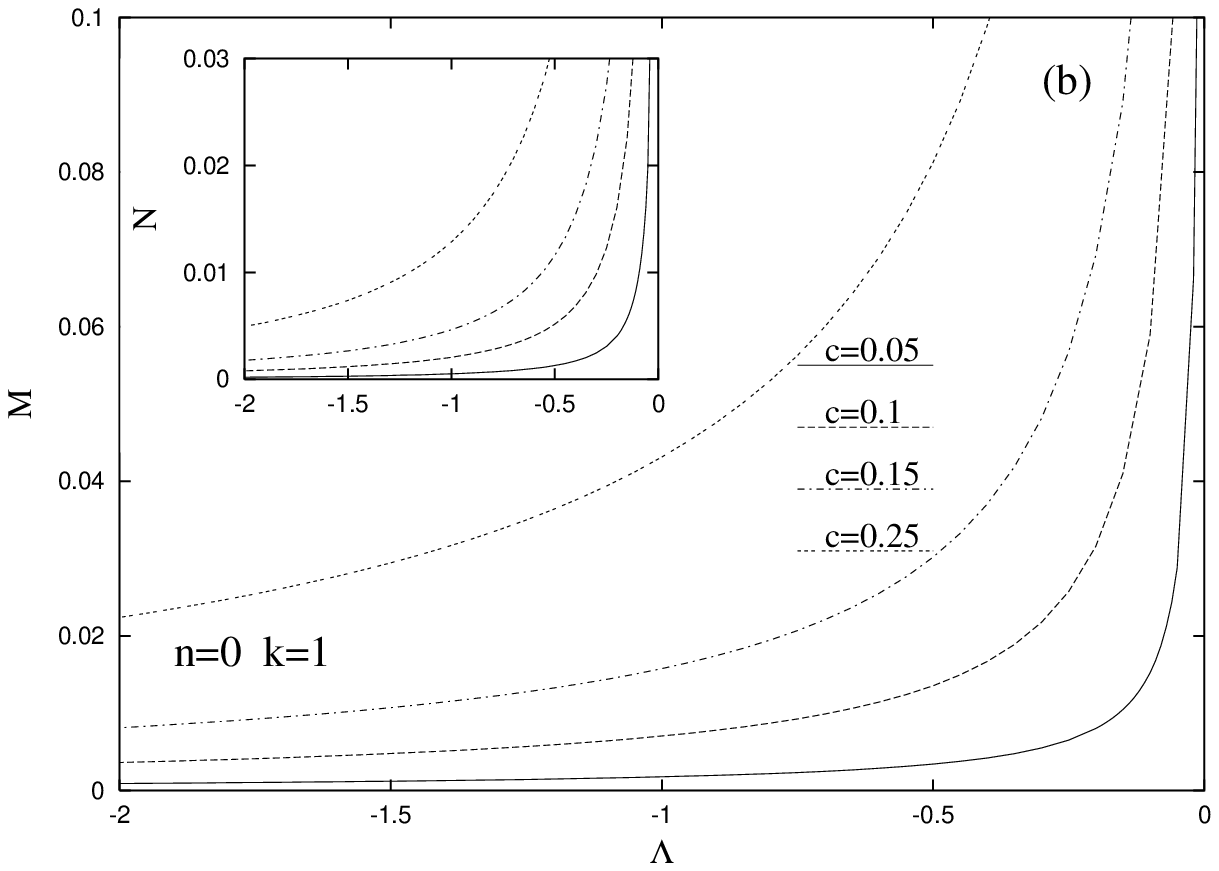,width=13cm}}
\end{picture}
\\
\\
\\
{\small{\bf  Figure 2.}  
The asymptotic value $M$ of the metric function  $m(r)$ 
and the particle number $N$ are represented 
as a function of the parameters $c$ and $\Lambda$ for rotating, 
nodeless solutions with 
azimuthal number $k=1$.}
\vspace{0.7cm} 
\\
resembles the large 
self-interaction BS configuration, where  
a vacuum hole at the center is always present \cite{Sakamoto:1999zt}.
Some properties of these solutions are better understood by studying 
the 
timelike and null geodesic motion.
The radial geodesic equation is
\begin{eqnarray}
\label{g1}
\dot{r}^2=-\varepsilon F+e^{2 \delta}(E+\Omega 
L)^2-\frac{F}{r^2}L^2=-V(r),
\end{eqnarray}
where a dot stands for a derivative with respect 
to the parameter $\tau$, and $\varepsilon=1$ or 
$0$ for timelike or null geodesics, respectively. 
$\tau$ is an affine parameter along the geodesics; for timelike 
geodesics, 
$\tau$ is the proper time.
$L$ and $E$ are two constants of motion associated with the Killing 
vectors
$\partial_{\varphi}$ and $\partial_{t}$, respectively.

The rotational motion $(L \neq 0)$ 
is determined by the centrifugal force term $FL^2/r^2$, which 
forbids a test particle to access the BS center.
Therefore any allowed rotational motion
has a minimal allowed value of the radial coordinate.
The radial motion $(L=0)$ of a massless test particle is unbounded, 
with 
 a speed $ dr/d\tau=E$ at spatial infinity, while $r=0$ is approached 
for some 
finite value of $\tau$.
The trajectory of massless particles with $L\neq 0$ is bounded for 
$L^2/l^2>E^2$ only.
All possible motions of a massive particle are bounded, since $V(r)$ 
is approximated for 
large $r$ as 
\begin{eqnarray}
\label{g2}
V(r)\sim  1-2M+\frac{r^2}{l^2}+\frac{L^2}{l^2}-E^2+O(1/r^2).
\end{eqnarray}
Massive particles with $L=0$ can approach the central 
region of the BS for $E^2>e^{-2\delta(0)}$ only.
\section{The mass and angular momentum}
The total mass $\bar{M}$ and angular momentum $\bar{J}$ of 
(2+1)-dimensional 
BS solutions with large self-interaction, 
discussed in refs.~\cite{Sakamoto:1998hq,Sakamoto:1999zt}, 
have been computed by using  a matching procedure 
on a surface $r=r_0$ separating the regions where 
the internal (BS region) and external geometries are defined. 
The external geometry (with a vanishing scalar field) 
is taken to be the BTZ black hole which gives 
the values of $\bar{M}$ and $\bar{J}$. 
However, in the study of numerical solutions it is desirable to 
avoid this method 
and, similar to the (3+1)-dimensional case, to consider 
a scalar field extending to infinity.

In computing quantities like mass and angular momentum one usually 
encounters
infrared divergences, which are regularized
by subtracting a suitably, chosen background
\cite{Gibbons:1976ue}. 
However, such a procedure does not work in general; 
in certain cases  the choice of reference background is 
ambiguous or unknown. In order to regularize such divergences, 
a different procedure has been proposed in 
ref.~\cite{Balasubramanian:1999re}.
This technique was inspired by the AdS/CFT correspondence and 
consists in adding suitable counterterms  $I_{ct}$  to the action. 
These counterterms are constructed from curvature invariants of the 
induced
boundary metric $h_{ab}$ --- they contribute an extra surface 
integral to the action and, obviously, the bulk equations of motion 
are
not altered. As originally found in 
ref.~\cite{Balasubramanian:1999re}, for 
vacuum solutions with a negative cosmological constant, the following 
counterterms are sufficient to cancel divergences in three dimensions
\begin{eqnarray}
\label{ct}
I_{\rm ct}=-\frac{1}{8 \pi G} \int_{\partial {\cal M}}d^{2}x\sqrt{-h}
\frac{1}{l}.
\end{eqnarray}
The boundary divergence-free stress tensor is given by
\begin{eqnarray}
\label{s1}
T_{ab}&=& \frac{2}{\sqrt{-h}} \frac{\delta I}{ \delta h^{ab}}
=\frac{1}{8\pi G}(K_{ab}-Kh_{ab}-\frac{1}{l}h_{ab}).
\end{eqnarray}
The efficiency of this approach has been demonstrated in a broad 
range of examples,
the counterterm subtraction method being developed on its own 
interest and applications.

If there are matter fields on $\cal{M}$, additional counterterms may 
be needed to 
regulate the action. 
In our case, since the scalar field behaves at infinity  
like $O(r^{-2-\epsilon})$, the 
counterterm given in eq. (\ref{ct}) is enough to yield a finite 
result for the 
boundary stress tensor (see, however, ref. \cite{Gegenberg:2003jr} 
for an 
example with a different asymptotic behavior,  where $\phi \sim O(r^{-1/2})$ and scalar field 
boundary counterterms $I_{ct}^{s}$ should be introduced in the action).
Using the asymptotic expressions (\ref{asymptotics}), we find
the following boundary stress tensor:
\begin{eqnarray}
\label{BD4}
8 \pi G T_{\varphi \varphi}=l(M-\frac{1}{2})+O(1/r),
~~
8 \pi GT_{\varphi t}=\frac{J}{l}+O(1/r),
~~
8 \pi G T_{tt}=\frac{1}{l}(M-\frac{1}{2})+O(1/r).
\end{eqnarray}
The conserved charges can be constructed by choosing an ADM foliation 
of $\partial M$, with spacelike surfaces $\Sigma$, so that
\begin{eqnarray}
\label{ADM}
h_{ab}dx^{a}dx^{b}=-N_{\Sigma}^2dt^2+\sigma (d 
\varphi+N_{\Sigma}^{\varphi}dt)^2.
\end{eqnarray}
In this approach, the conserved quantities associated with a Killing 
vector 
of the boundary $\xi^{b}$, are given by
\begin{eqnarray}
\label{charge}
Q(\xi)=\int_{\Sigma}d\varphi \sqrt{\sigma} u^{a}T_{ab}\xi^{b},
\end{eqnarray}
where $u^{b}$ is the unit timelike vector normal to $\Sigma$. 
The result we find for the mass  is $\bar{M}=(2M-1)/8G$ and
is always larger than $-1/8G$, with the extreme value 
$\bar{M}=-1/8G$ corresponding to the global $AdS_3$ space. Contrary 
to 
what happens for BTZ black holes, we do not find a mass gap between 
$\bar{M}=-1/8G$ and $\bar{M}=0$. 
The total angular momentum $\bar{J}$ 
of these rotating BS solutions is the charge associated with 
the Killing vector $\partial_{\varphi}$ and has the value $J/4G$.

\section{Remarks on charged rotating boson stars}
The (2+1)-dimensional case is also an ideal arena to  
study more complicated rotating configurations, 
which are very difficult to discuss in higher dimensions.
An obvious example is to include a $U(1)$ gauge field in the action 
principle 
and to look for charged rotating BS (the static, four-dimensional 
counterparts
of these configurations
are found in ref.~\cite{Jetzer:1989av}).
Here, the 
total angular momentum
includes a supplementary gauge field contribution
which may lead to a violation of the generic relation (\ref{JN}).  
The electromagnetic field lagrangean
$L_{em}=-1/4F_{ij}F^{ij}\sqrt{-g}$, 
(with $F_{ij}=\partial_{i}A_{j}-\partial_{j}A_{i}$),
is introduced in the action principle (\ref{action}),
while in the general expresions involving the scalar field, 
the ordinary derivative
is replaced by gauge derivative 
$\partial_{j}\Phi \to D_{j}\Phi=\partial_{j}\Phi+i e A_{j} \Phi,$ 
$e$ being the gauge coupling constant.
The electromagnetic field equations read
\begin{eqnarray}
\label{maxwell}
\nabla_{j}F^{j l}=e J^{l},
\end{eqnarray}
while the expression (\ref{Tij}) of the energy-momentum tensor  includes
a supplementary $U(1)$ contribution
\begin{eqnarray}
\label{Tijs}
T_{ij}=D_{i}\Phi^{\ast}D_{j}\Phi +D_{j}\Phi^{\ast} D_{i}\Phi  
-g_{ij}\left(g^{lm}D_{l}\Phi^{\ast} D_{m}\Phi+V(|\Phi|^2)\right)
+F_{il}F_{jm}g^{lm}-\frac{1}{4}g_{ij}F_{lm}F^{lm}.
\end{eqnarray}
Here we notice the existence of an
intriguing expression for the volume integral (\ref{J}), the 
contribution of the electromagnetic field to the total angular momentum
admitting a representation as a surface integral at infinity.
For the same form of the scalar field and an $U(1)$ field ansatz
\begin{eqnarray}
\label{em}
A=A_{\varphi} d \varphi+A_t dt,
\end{eqnarray}
the  component of the
energy-momentum tensor asssociated with rotation reads
\begin{eqnarray}
\label{T34}
T_{\varphi}^t=
D_{\varphi}\Phi^{\ast}D^{t}\Phi+
D_{\varphi}\Phi D^{t}\Phi^{\ast}+F_{\varphi l}F^{t l}=
kJ^t+\frac{1}{\sqrt{-g}}\partial_{j}\left(A_{\varphi}F^{j t}\sqrt{-g}\right),
\end{eqnarray}
$k$ being again the
vorticity number and we used also the Maxwell equations (\ref{maxwell}).
This leads to the simple result
\begin{eqnarray}
\label{Jd}
\int T_{\varphi}^t \sqrt{-g}=kN+\oint_{\infty}A_{\varphi} F^{jt}dS_j,
\end{eqnarray}
where $N$ is the particle number.\footnote{Note also that the equations (\ref{maxwell})
implies the relation $Q_e=e N$ between the electric charge of solutions and the particle number.}
However, different form the pure scalar field case, 
this integral cannot be identified with the
(finite-)total angular momentum.
The charged rotating BS solutions asymptotically approach,
to leading order, the rotating $U(1)$ BTZ 
solutions discussed in \cite{Clement:1993kc, Martinez:1999qi}.
The solution of the field equations (\ref{maxwell})
as $r \to \infty $ is $A_t\sim Q_e\log(r),~A_{\varphi}\sim C\log(r)$ 
which modify the asymptotic behavior 
of the metric functions $m(r)$ and $\Omega(r)$, leading to logarithmic terms
(the scalar field still decays as $r^{-2-\epsilon}$).
The presence of divergences at spatial infinity in the mass and angular momentum is 
an usual feature in electrically charged $AdS_3$ solutions.
The mass and angular momentum of charged rotating BS
configurations can be computed by using the methods developed 
for rotating $U(1)$ BTZ case 
\cite{ Martinez:1999qi, Clement:2003sr}.
Here we note that the counterterm formalism does not work in this case,
and it is necessary to include
a supplementary $I_{ct}^{em}$ term in the general action.

Numerical solutions describing charged rotating BS 
can easily be obtained by using the same approach
presented above.
We have integrated the field equations for several values of the cosmological constant,
looking for configurations approaching at infinity the charged rotating BTZ background. 
Static charged solutions with $k=A_{\varphi}=0$ have been found as well.
However, the presence of an electromagnetic field complicates very much the general picture
(for example, there are two possible sets of boundary conditions at $r=0$).
The solution properties depend also on the value of the coupling constant $e$ and differ
significantly from the known four dimensional static charged boson stars.
A study of these solutions, as well as a computation of their mass and 
angular momentum will be presented 
elsewhere.

The result (\ref{Jd}) can easily be generalized to higher dimensions, 
for values of the cosmological constant $\Lambda \leq 0$ and any scalar field potential 
(it holds also in the absence of gravity as well).
In the  more interesting four dimensional case, 
we expect to find asymptotically a rotating
Kerr-Newmann-(AdS) configuration, 
with $A_{\varphi} \sim P \cos \theta,~~F^{rt} \sim Q/r^2$ 
($P$ and $Q$ corresponding to magnetical, respectiv electrical charge) 
which yields a finite angular momentum 
\begin{eqnarray}
\label{J4}
\bar{J}=kN.
\end{eqnarray}
Thus, the full angular momentum
is carried by the complex scalar field, and, similar 
to the nonabelian dyons \cite{VanderBij:2001nm, vanderBij:2002sq}, the total
angular momentum of the gauge field is zero.
The construction of such charged rotating four dimensional BS configurations represents a
difficult challenge. 
%
\section{Conclusions and discussion}
We have presented arguments that a gravitating complex scalar field 
model in (2+1) dimensions admits, in the case of a negative cosmological 
constant,  a continuous family of regular rotating solutions.
Here, numerical solutions are already known in the literature 
in the presence of a self-interaction term $\lambda |\Phi|^4$, as
$\lambda \to \infty$ \cite{Sakamoto:1999zt}.
However, this limit makes obscure the influence 
of a number of physical parameters on the solutions 
properties, and imposes a rather unnatural assumption of scalar field
confinement in a finite region of spacetime.

The configurations we have found can be regarded as the 
lower-dimensional
counterparts of the well-known rotating BS solutions in 
(3+1)-dimensions \cite{Yoshida:1997qf,Schunck}, presenting a number 
of common properties. 
However, they are asymptotically AdS, 
whereas, without self-interaction, no regular solutions are found in the limit 
$\Lambda \to 0$.  The cosmological constant acts here as an 
attractive gravitational force, increasing with the radial distance,
that is balanced by the scalar field pressure and the centrifugal 
force.
These solutions can be labeled by $(\phi^{(|k|)}(0),n,k)$, where 
$\phi^{(p)}(0)$ is the $p$-th derivative of the scalar field 
evaluated at 
the origin, $n$ is the node number of the scalar field and
$k$ is  the vorticity number. The solutions with $k \neq 0$ can be 
regarded as spinning excitations of the fundamental $k=0$ static 
solutions.
Similar to other known rotating solitons, 
their angular momentum is uniquely determined by the value of 
particle number.

To address the question of stability\footnote{
A study of stability of a $D-$dimensional static BS solutions with 
$\Lambda \leq 0$ under 
linear perturbations was presented in ref.~\cite{Astefanesei:qy},
leading to a Sturm--Liouville-type eingenvalue problem.
However, the significance of these results in (2+1) dimensions is 
unclear.}, we consider
the binding energy of these BS solutions, whose natural definition in
(2+1) dimensions is \cite{Sakamoto:1998hq}
\begin{eqnarray}
\label{bi} 
E_b \equiv \bar{M}+\frac{1}{8 G}-\mu N.
\end{eqnarray}
Similar to higher dimensions, this quantity must be negative for 
stable configurations.
The situation we find here is more complicated, depending on the 
value of cosmological constant.
For small values of $|\Lambda|$, $E_b$ is negative for all allowed 
values of $c$.
Once $|\Lambda|$ is increased, we notice the $E_b<0$ for $c>c_0$ 
only, while $c_0 \to c_{max}$
for large enough  $|\Lambda|$ 
(for example, the $\Lambda=-1$ configurations presented in Fig.2(a) 
have $E_b<0$, 
those with $\Lambda=-4$ have positive binding energy, while the 
$\Lambda=-2$ solutions
with $c>0.95$ have $E_b<0$).

We expect to obtain a very similar qualitative behavior of the 
solutions
when discussing a number of possible extensions of this theory, 
$e.g.$ including 
a dilaton or a nonminimal coupling term $\xi \Phi ^{\ast} \Phi$R term 
between the
scalar field and gravity.

Since a complex scalar field is present in many supergravity 
theories, 
one may ask about the possible relevance of 
BS solutions within the holographic principle and its AdS/CFT 
correspondence realization.
A scalar field has been discussed in this context by many authors, 
however without considering this type of 
"{\it macroscopic quantum states}" \cite{Mielke:1997re}. 

The (2+1)-dimensional case is rather special, since we could have 
renormalizable 
pure Einstein gravity on AdS{$_3$} that can be written as a 
Chern-Simons theory.
The background metric upon which the dual field theory resides is 
just 
the rescaled boundary metric
\begin{eqnarray}
\gamma_{ab}=\lim_{r \rightarrow \infty} \frac{l^2}{r^2}h_{ab},
\end{eqnarray}
giving the line element
$\gamma_{ab}dx^a dx^b=-dt^2+l^2d\varphi^2,$ which, for a BS solution 
corresponds to a static cylinder 
(note also that for a rotating black 
hole in the bulk, the boundary is rotating \cite{Hawking:1998kw}).
There is now considerable evidence that the 
boundary conformal field theory corresponding to the bulk 
Chern-Simons 
theory is a Liouville field theory (see, $e.g.$, 
ref.~\cite{henneaux}). 
After canonical quantization, the spectrum of Liouville theory 
comprises of two 
different classes \cite{seiberg}: macroscopic (normalizable) states 
and microscopic (non-normalizable) states. 
The normalizable states of Liouville theory give a CFT with a 
central charge $c_{eff}=1$ (see, $e.g.$, 
refs.~\cite{Martinec:1998wm,Myung:1998jw}) 
and so, one expects bulk solutions without horizon would 
correspond 
to macroscopic Liouville states.\footnote{A recent proposal to 
explain BTZ black hole entropy by considering the non-normalizble modes in 
the boundary Liouville theory can be found in ref.~\cite{Chen:2003si}.} 

The presence of additional matter fields in the bulk will not 
qualitatively change this picture: the Virasoro algebra is an asymptotic 
isometry of $AdS_3$ but the boundary theory is not Liouville theory. 
However, more generally, one can conjecture that any consistent 
quantum gravity on $AdS_3$ is dual to a two-dimensional CFT living on the 
boundary.
In light of the AdS/CFT correspondence,  
Balasubramanian and Kraus have interpreted eq.~(\ref{s1}) as giving 
the
expectation value of the dual theory stress tensor 
$<\tau^{ab}>=\frac{2}{\sqrt{-\gamma}} \frac{\delta S_{eff}}{\delta 
\gamma_{ab}}$.
The relation between $<\tau^{ab}>$ and the boundary stress-tensor is 
\cite{Myers:1999qn}
\begin{eqnarray}
\label{r1}
\sqrt{-\gamma}\gamma^{ab}<\tau_{bc}>=
\lim_{r \rightarrow \infty} \sqrt{-h} h^{ab}T_{bc}.
\end{eqnarray}
 Different bulk configurations can have the same asymptotia 
(are locally AdS) and so the 
expectation value of the  dual CFT stress tensor is insufficient 
to distinguish between them (a leading order expression similar 
to eq.~(\ref{BD4}) is found also for a BTZ black hole). However, the 
obstructions of extending the solutions in the bulk will be resolved 
by additional CFT data. 

Here we remark that a scalar field in the bulk has both kind of 
modes: 
normalizable and non-normalizable. 
It is clear now that boundary conditions at infinity and 
initial conditions for the bulk fields are essentially in this 
picture. Sources in the dual CFT determine an asymptotic expansion of the 
corresponding field near the boundary, in other words the 
non-normalizable bulk modes are equivalent with local operator insertions on the 
boundary. 
On the other hand, the normalizable modes are fluctuating in the bulk 
(for fixed boundary conditions) and quanta occupying such modes have 
a dual description in the boundary Hilbert space \cite{bala1,bala2}. 
Boson star is an example of soliton whose field depends on time and 
does not have a non-trivial topology: the existence and stability of 
the soliton are governed by its charge, $N$ --- in the language of 
particles, there would have to be $N$ charged particles. In analogy with 
the pure gravity case, it would be interesting to find an 
interpretation of boson stars (macroscopic quantum states without horizon) as 
dual macroscopic (zero temperature) CFT states described by acting on 
the vacuum with modes of the appropiate boundary operator 
\cite{bala2}.

\subsection*{Acknowledgments}
The authors are grateful to 
Yoonbai Kim for interesting remarks on a first version of this paper.
DA would like to thank Kiril Krasnov for interesting conversations.
 DA also thanks the 
University of Waterloo's Department of Physics
for their ongoing hospitality during this project. 

DA is supported by a Dow Hickson Fellowship. 
The work of ER was performed in the context of the Graduiertenkolleg 
of the Deutsche Forschungsgemeinschaft (DFG): Nichtlineare 
Differentialgleichungen: Modellierung,
Theorie, Numerik, Visualisierung.


\end{document}